\documentclass[aps,twocolumn,epsfig,rotate]{revtex4}
\usepackage{epsfig}
\usepackage{graphicx}
\usepackage{amsmath}

\newcommand{\beq}{\begin{eqnarray}}
\newcommand{\eeq}{\end{eqnarray}}
\begin{document}                
\title{Measurement-assisted Coherent Control}
\author{Jiangbin Gong and Stuart A. Rice}
\affiliation{Department of Chemistry and The James Franck Institute,\\
The University of Chicago, Chicago, Illinois 60637}
\date{\today}

\begin{abstract}
Two advantageous roles of the influence of measurement on a system subject to coherent  control
are exposed using a five-level
model system.  In particular,
a continuous measurement of the population in a
branch state in 
the Kobrak-Rice extended stimulated Raman adiabatic passage scheme
is shown to provide a powerful means for controlling
the population transfer branching
ratio between two degenerate target states. It is demonstrated that 
a measurement with a large strength
may  be used to
completely shut off the yield of one target state
and that the same measurement with a weak strength
can dramatically enhance the robustness of the controlled 
branching ratio against dephasing.
\end{abstract}
\pacs{32.80.Qk, 03.65.Xp}
\maketitle

\section{Introduction}
Interest in coherent control of atomic and molecular processes 
\cite{ricebook,rice01,brumerbook,bergmann,rabitz} has grown rapidly 
in recent years.  One of the prominent goals is the manipulation of quantum interference
effects to enhance the yield of a particular wanted final state
while suppressing the yield of another. An example of this class of operations 
is enhancement of the yield of one product of a reaction while suppressing the yield of
a competing product. In this case, and in others as well, 
a quantum system under control is typically not measured {\it until} the desired
unitary evolution ends.  
The advantages and disadvantages of the influence of measurement on the system
in a variety of coherent control scenarios remain unexplored.

We note that the inhibition or acceleration of a
quantum process under observation, 
the so-called quantum Zeno and anti-Zeno effects,
demonstrate that the role
of measurement in quantum dynamics can be dramatic.  
Although the quantum Zeno effect was first derived in terms of
wavefunction collapse induced by a measurement, recent studies \cite{vera,schulman,luis,facchi2}
have shown that
introducing the concept of wavefunction collapse is unnecessary for describing the quantum Zeno and anti-Zeno effects. 
That is, a measurement apparatus, even in the case of obtaining a null result,
always disturbs the system's Hamiltonian
and can therefore
strongly affect the quantum dynamics without involving our consciousness,
a view that is adopted in this paper. 

In this paper we describe two
interesting and advantageous roles of measurement in coherent control,
using a well-studied model of selective photochemistry, namely,
the Kobrak-Rice extended five-level stimulated Raman adiabatic passage scheme \cite{bergmann,kobrakpra,kurkal}. That model is
also the basis for the Chen-Shapiro-Brumer strong field control approach to selective photochemistry \cite{brumerbook}.
In particular, it is shown that
a continuous measurement of population in a so-called ``branch state" with a large measurement strength
provides a powerful means for manipulating the nonadiabatic coupling
and therefore the useful quantum interference
between two particular adiabatic states. The resultant 
control over the population transfer branching
ratio between two degenerate product states is extraordinary insofar as 
(1) the yield of one of the two target states may be completely shut off while the
yield of the other target state is still considerable, and (2) the intermediate state is
unpopulated during the population transfer.
Further, we
show that
the same measurement with a weak strength
can dramatically enhance the stability of the controlled
branching ratio under adverse situations, such as in the presence of strong dephasing modelled by stochastic energy level
fluctuations.
These results are of broad theoretical and experimental interest. Indeed, the measurement-assisted
coherent control of the branching ratio between two unmeasured product states 
can be
regarded as a third aspect of measurement effects that complements the
quantum Zeno and anti-Zeno effects.

\begin{figure}[ht]
\begin{center}
\epsfig{file=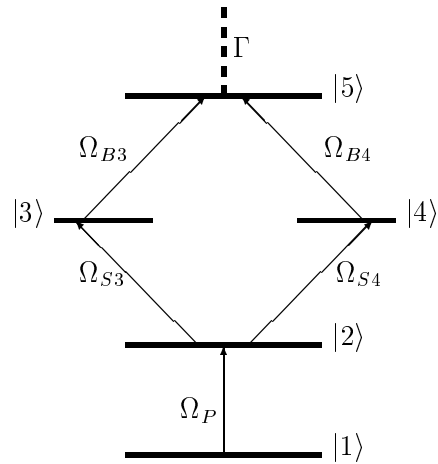,width=6.5cm}
\end{center}
\vspace{-1cm}
\caption{A schematic diagram of the Kobrak-Rice five-level model \cite{kobrakpra} for the control of population
transfer branching ratio between two degenerate states $|3\rangle$ and  $|4\rangle$,
with the branch state $|5\rangle$
subject to a continuous population measurement.}
\label{fig1}
\end{figure}

\section{Exploiting Measurement Effects: Strong measurement Limit}
A schematic diagram 
of the Kobrak-Rice five-level model is shown
in Fig. \ref{fig1}. The initial state $|1\rangle$ is coupled with the intermediate state $|2\rangle$ by a pump pulse,
while state $|2\rangle$ is coupled with two degenerate product states $|3\rangle$ and $|4\rangle$ by a Stokes pulse.
Due to the  degeneracy of states $|3\rangle$ and $|4\rangle$, altering their population transfer branching ratio (denoted by $B$)
cannot be achieved
unless we introduce
the ``branch state" $|5\rangle$ that is also coupled with states $|3\rangle$ and $|4\rangle$ by a third ``branching
pulse''.
All the laser fields are on resonance with the respective transitions \cite{note2} and are assumed to be Gaussian, with their pulse
duration sufficiently short that, except for one case discussed below,
the natural lifetimes of all the five states can be taken as infinity. 
To be specific,
the associated 
Rabi frequencies (assumed to be real) [see  Fig. \ref{fig1}] are chosen to be
$\Omega_{P}=\tilde{\Omega}_{P}\exp[-(t-T)^{2}/T^{2}]$,
$\Omega_{S3,S4}=\tilde{\Omega}_{S3,S4}\exp[-t^{2}/T^{2}]$,
and $\Omega_{B3,B4}=\tilde{\Omega}_{B3,B4}\exp[-0.5(t-0.5T)^{2}/T^{2}]$, where the pump and Stokes pulses 
have been counter-intuitively ordered \cite{bergmann}. The characteristic magnitude of the various peak Rabi frequencies is
represented by $\tilde{\Omega}$. 
The branch state is also subject to a continuous population measurement by coupling it to a continuum.
Within the framework of the slowly varying continuum approximation \cite{brumerbook} the disturbance of the system by
the continuous measurement
can be described by a finite and controllable lifetime of state $|5\rangle$ \cite{schulman,facchi2}, 
with a decay rate constant (denoted $\Gamma$) to be tuned via
the strength of the system-probe interaction.  
Using the rotating wave approximation and the interaction representation,
the system Hamiltonian is then given by
$H=H_{r}+H_{\Gamma}$,
where
$H_{r}$ denotes the on-resonance Hamiltonian in the absence of branch state measurement,
\begin{equation}
H_{r}=\left[
\begin{array}{ccccc}
0 & \Omega_{P} & 0  & 0 &  0 \\
\Omega_{P} & 0 & \Omega_{S3} &  \Omega_{S4} & 0 \\
0 & \Omega_{S3} & 0 & 0 & \Omega_{B3} \\
0 &  \Omega_{S4}& 0 &  0 & \Omega_{B4} \\
0 & 0 & \Omega_{B3} & \Omega_{B4} & 0\\
\end{array}
\right],
\end{equation}
and $\langle k |H_{\Gamma}|j\rangle=-i\Gamma \delta_{k5}\delta_{j5}$ is due to
the continuous measurement. 
The eigenvalues and eigenvectors of $H_{r}$, denoted by
$\lambda_{k}$ and $|\lambda_{k}\rangle$, $k=1-5$, can be obtained analytically. In particular,
there exists one null eigenvalue (denoted $\lambda_{1}$) with its eigenvector given by
\beq
|\lambda_{1}\rangle=\frac{1}{N_{1}}\left[\Omega_{SB},0, -\Omega_{P}\Omega_{B4}, \Omega_{P}\Omega_{B3},0\right]^{T}, 
\label{eq1}
\eeq
where $\Omega_{SB}\equiv (\Omega_{S3}\Omega_{B4}-\Omega_{S4}\Omega_{B3})$.
Other eigenvalues of $H_{r}$ are given by
\beq
\lambda_{k}^{2}=\frac{1}{2}\left(\Omega^{2}_{M}\pm\sqrt{\Omega^{4}_{M}-4[\Omega_{SB}^{2}+
\Omega_{P}^{2}(\Omega_{B3}^{2}+\Omega_{B4}^{2})]}\right), 
\eeq
where $k=2-5$, and $\Omega_{M}^{2}$ represents the sum of the squares of all the five Rabi frequencies.
The associated eigenvectors are given by
\beq
|\lambda_{k}\rangle=
\frac{1}{N_{k}}\left[
\begin{array}{c}
\Omega_{P}(\lambda_{k}^{2}-\Omega_{B3}^{2}-\Omega_{B4}^{2})\\
\lambda_{k}(\lambda_{k}^{2}-\Omega_{B3}^{2}-\Omega_{B4}^{2})\\
\Omega_{S3}\lambda_{k}^{2}-\Omega_{B4}\Omega_{SB}\\
\Omega_{S4}\lambda_{k}^{2}+\Omega_{B3}\Omega_{SB}\\
\lambda_{k}(\Omega_{B3}\Omega_{S3}+\Omega_{B4}\Omega_{S4})\\
\end{array} 
\right], \ k=2-5,
\label{eigenk}
\eeq
where $N_{k}$ is the normalization factor.
Note that $|\lambda_{1}\rangle$ has a node on state $|2\rangle$, can fully
correlate with the initial and final states, and can be far away from $|\lambda_{k}\rangle$ ($k=2-5$).
In the absence of $H_{\Gamma}$, 
$|\lambda_{1}\rangle$ is the key eigenstate 
for adiabatic passage \cite{bergmann} without populating the intermediate state,
and the associated population transfer branching ratio between the two degenerate target states $|3\rangle$ and $|4\rangle$  is given by
$B=B_{1}\equiv \tilde{\Omega}_{B4}^{2}/\tilde{\Omega}_{B3}^{2}$.  
Interestingly, $|\lambda_{1}\rangle$ [see Eq. (\ref{eq1})] also has zero overlap with the branch state, so it survives
in the presence of $H_{\Gamma}$. However, the exact forms of the other eigenvalues and eigenvectors of $H$  
are too complicated to be useful.  Below we use various perturbation theory techniques to reveal the interesting
consequences for the system of introducing $H_{\Gamma}$.

Here we consider  the case of strong system-probe interaction, {\it i.e.}, $\Gamma>>\tilde{\Omega}$.
Then the system is dominated by the measurement and
the laser-induced couplings can be regarded as a perturbation.   In particular,
$H$ can be written as $H=(H-V_{B})+V_{B}$, where 
\beq
\langle k|V_{B}|j\rangle\equiv
\Omega_{B3}(\delta_{k3}\delta_{j5}+\delta_{j3}\delta_{k5})+
\Omega_{B4}(\delta_{k4}\delta_{j5}
+\delta_{j4}\delta_{k5}).
\eeq
The eigenvalue-eigenvector structure of $(H-V_{B})$ can be easily obtained since state $|5\rangle$ therein is decoupled by construction.
However, because
$(H-V_{B})$ has two (degenerate) null eigenvalues, a naive perturbation treatment is bound to fail.
Our strategy is to first  make a special superposition of the two null eigenstates of $(H-V_{B})$
to construct the eigenstate $|\lambda_{1}\rangle$ that is known to exist in the presence of $V_{B}$.
Then second-order perturbation theory is applied to the subspace that is 
orthogonal to $|\lambda_{1}\rangle$.
We obtain three complex eigenvalues and one purely imaginary eigenvalue (denoted $\lambda_{2}'$)
that lies closest to $\lambda_{1}$. As such, for the consideration of nonadiabatic effects below,  only the
coupling between $|\lambda_{1}\rangle$ and the eigenvector  
associated with $\lambda_{2}'$  (denoted $|\lambda_{2}'\rangle$) is important.
Specifically,
\beq
\lambda_{2}'=-i\left[\Omega_{P}^{2}(\Omega_{B3}^{2}+\Omega_{B4}^{2})+
\Omega_{SB}^{2}\right]^{2}/[\Gamma (N_{2}')^{2}];
\label{l2}
\eeq
\beq
|\lambda_{2}'\rangle=\frac{1}{N_{2}'}
\left[\begin{array}{c}
\Omega_{P}(\Omega_{S4}\Omega_{B4}+\Omega_{S3}\Omega_{B3})\\
0\\
\Omega_{S3}\Omega_{S4}\Omega_{B4}-\Omega_{B3}(\Omega_{P}^{2}+\Omega_{S4}^{2})\\
\Omega_{S3}\Omega_{S4}\Omega_{B3}-\Omega_{B4}(\Omega_{P}^{2}+\Omega_{S3}^{2}) \\
i\left[\Omega_{P}^{2}(\Omega_{B3}^{2}+\Omega_{B4}^{2})+\Omega_{SB}^{2}\right]/
\Gamma\\
\end{array}\right],
\label{eq2}
\eeq
where 
$N_{2}'$ is the normalization factor.
Equation (\ref{l2}) shows that the peak value of $i\lambda_{2}'$ is of the order of $\tilde{\Omega}^{2}/\Gamma$.
Hence, in effect, the continuous measurement of the population of the branch state with a large strength modifies the spectrum of $H_{r}$ such that 
an eigenstate ($|\lambda_{2}'\rangle$) 
that lies close to
$|\lambda_{1}\rangle$ is created and 
the spacing between them ($|\lambda_{1}-\lambda_{2}'|$) becomes tunable.
Equation (\ref{eq2}) shows that $|\lambda_{2}'\rangle$ also has a node on the intermediate state
but does not
overlap with the initial state at early times, and 
that the (final) $B$ associated with $|\lambda_{2}'\rangle$ is given by
$B_{2}\equiv\tilde{\Omega}_{B3}^{2}/\tilde{\Omega}_{B4}^{2}
=1/B_{1}$.


Let $\tilde{C}_{1}(\xi) $ and $\tilde{C}_{2}(\xi)$ represent the quantum amplitudes on the adiabatic states
$|\lambda_{1}\rangle$ and $|\lambda_{2}'\rangle$, where $\xi\equiv t/T$.
With all other eigenstates of $H$ neglected
the quantum dynamics reduces to that of a two-level system,
\beq
d\tilde{C}_{1}(\xi)/d\xi&=&O_{21}(\xi)\tilde{C}_{2}(\xi),
\label{eq6} \\
d\tilde{C}_{2}(\xi)/d\xi&=&-i\lambda_{2}'T\tilde{C}_{2}(\xi)-O_{21}(\xi)\tilde{C}_{1}(\xi),
\label{eq7}
\eeq
where $O_{21}(\xi)=\langle\lambda_{1}|d\lambda_{2}'/d\xi\rangle$ describes the nonadiabatic coupling
between $|\lambda_{1}\rangle$ and $|\lambda_{2}'\rangle$. Using Eqs. (\ref{eq1}) and (\ref{eq2}) one finds
\beq
O_{21}(\xi)=-2\Omega_{P}\Omega_{SB}(\Omega_{S3}\Omega_{B3}+\Omega_{S4}\Omega_{B4})/(N_{1}N_{2}').
\label{o21}
\eeq

Consider then three different regimes of $\Gamma$. First, in the adiabatic limit, {\it i.e.},
 $(\tilde{\Omega}T)^{2}>>\Gamma T$,
the decay term $-i\lambda_{2}'T$ in Eq. (\ref{eq7}) dominates such that
$\tilde{C}_{2}$ is always negligible and so is the population loss from
$|\lambda_{1}\rangle$. This gives $B\approx B_{1}$. Second,
in the quantum Zeno limit, {\it i.e.}, $(\tilde{\Omega}T)^{2}<<\Gamma T$,
$-i\lambda_{2}'T$ is negligible and $|\lambda_{1}\rangle$
and $|\lambda_{2}'\rangle$ become essentially degenerate, resulting in strong nonadiabatic effects that are beyond our control.
Indeed, in this case the branch state is observed too vigorously:
any transition to it is frozen and so it becomes useless for modulating control. The associated
$B$ then approaches $\tilde{\Omega}_{S3}^{2}/\tilde{\Omega}_{S4}^{2}$, a result inherent to the four-level system
obtained by discarding state $|5\rangle$.  The third regime, {\it i.e.}, $(\tilde{\Omega}T)^{2} \sim
\Gamma T$, is of most significance.  Denoting 
$\tilde{C}_{1}$ and $\tilde{C}_{2}$ as the final values of
$\tilde{C}_{1}(\xi)$ and $\tilde{C}_{2}(\xi)$ obtained from Eqs. (\ref{l2}), (\ref{eq6}), (\ref{eq7}) and (\ref{o21}),
and using Eqs. (\ref{eq1}) and (\ref{eq2})
one obtains
\beq
B=\frac{|\tilde{C}_{1}|^{2}\tilde{\Omega}_{B4}^{2}+|\tilde{C}_{2}|^{2}\tilde{\Omega}_{B3}^{2}
        +(\tilde{C}_{1}\tilde{C}_{2}^{*}+\tilde{C}_{1}^{*}\tilde{C}_{2})
        \tilde{\Omega}_{B3}\tilde{\Omega}_{B4}}
{|\tilde{C}_{1}|^{2}\tilde{\Omega}_{B3}^{2}+|\tilde{C}_{2}|^{2}\tilde{\Omega}_{B4}^{2}
        -(\tilde{C}_{1}\tilde{C}_{2}^{*}+\tilde{C}_{1}^{*}\tilde{C}_{2})
        \tilde{\Omega}_{B3}\tilde{\Omega}_{B4}}.
\label{key1}
\eeq
Clearly, the value of $B$ predicted by Eq. (\ref{key1}) is 
not a simple mixture of the two branching ratios $B_{1}$ and $B_{2}$ associated
with states $|\lambda\rangle$ and $|\lambda_{2}'\rangle$.
Rather, due to their constructive or destructive quantum  interference
$B$ 
can be zero or infinity. That is, by varying $\Gamma$ and therefore $\tilde{C}_{1}$ and  $\tilde{C}_{2}$,
different superpositions of the two adiabatic states $|\lambda_{1}\rangle$ and  $|\lambda_{2}'\rangle$
can be realized and
it becomes possible to completely suppress population transfer to one of the two
degenerate target states.
Detailed
conditions for $B=0$ or $B=\infty$ will be discussed elsewhere \cite{gong}.

\section{Examples}
\begin{figure}[ht]
\begin{center}

\epsfig{file=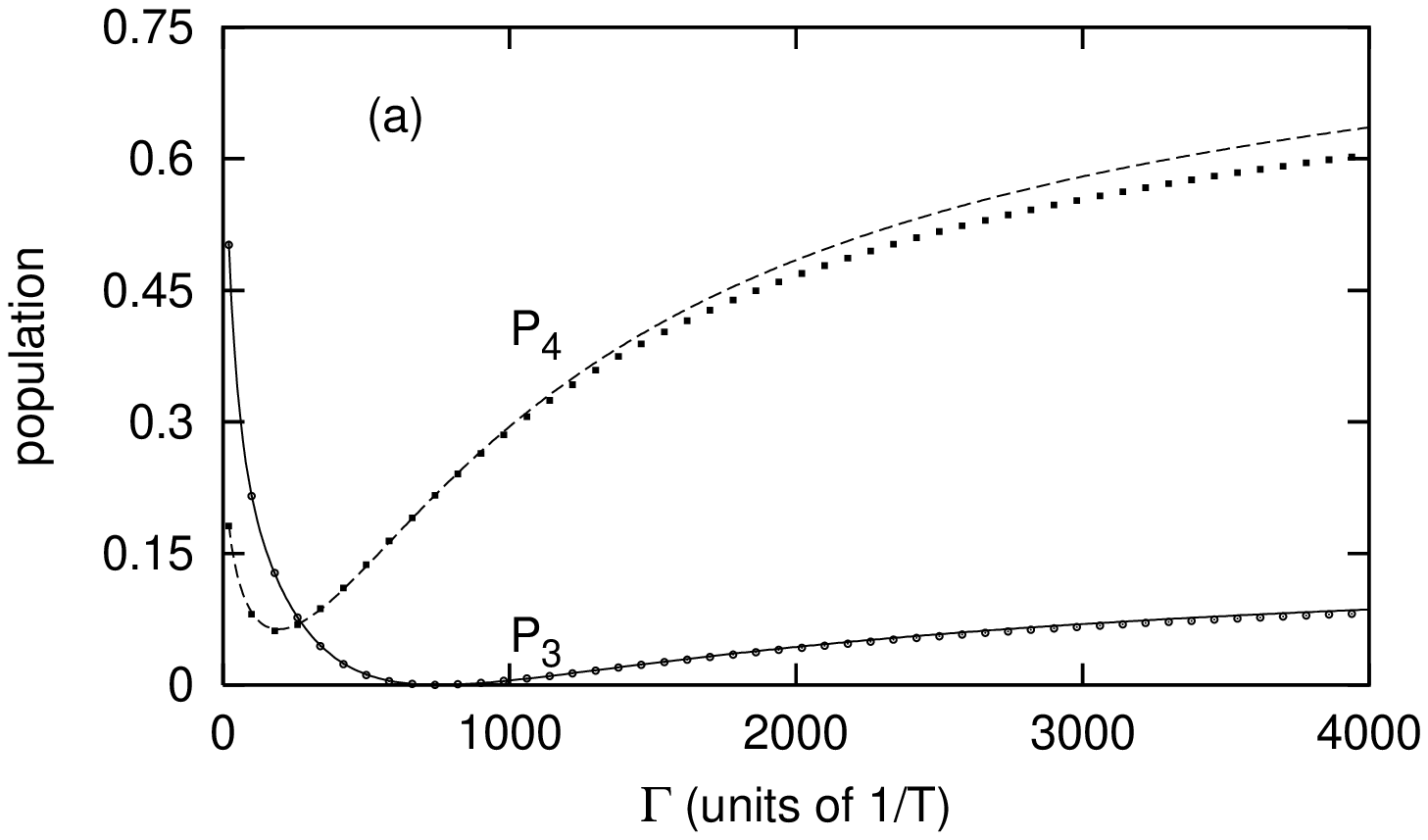,angle=270,width=6.5cm}

\epsfig{file=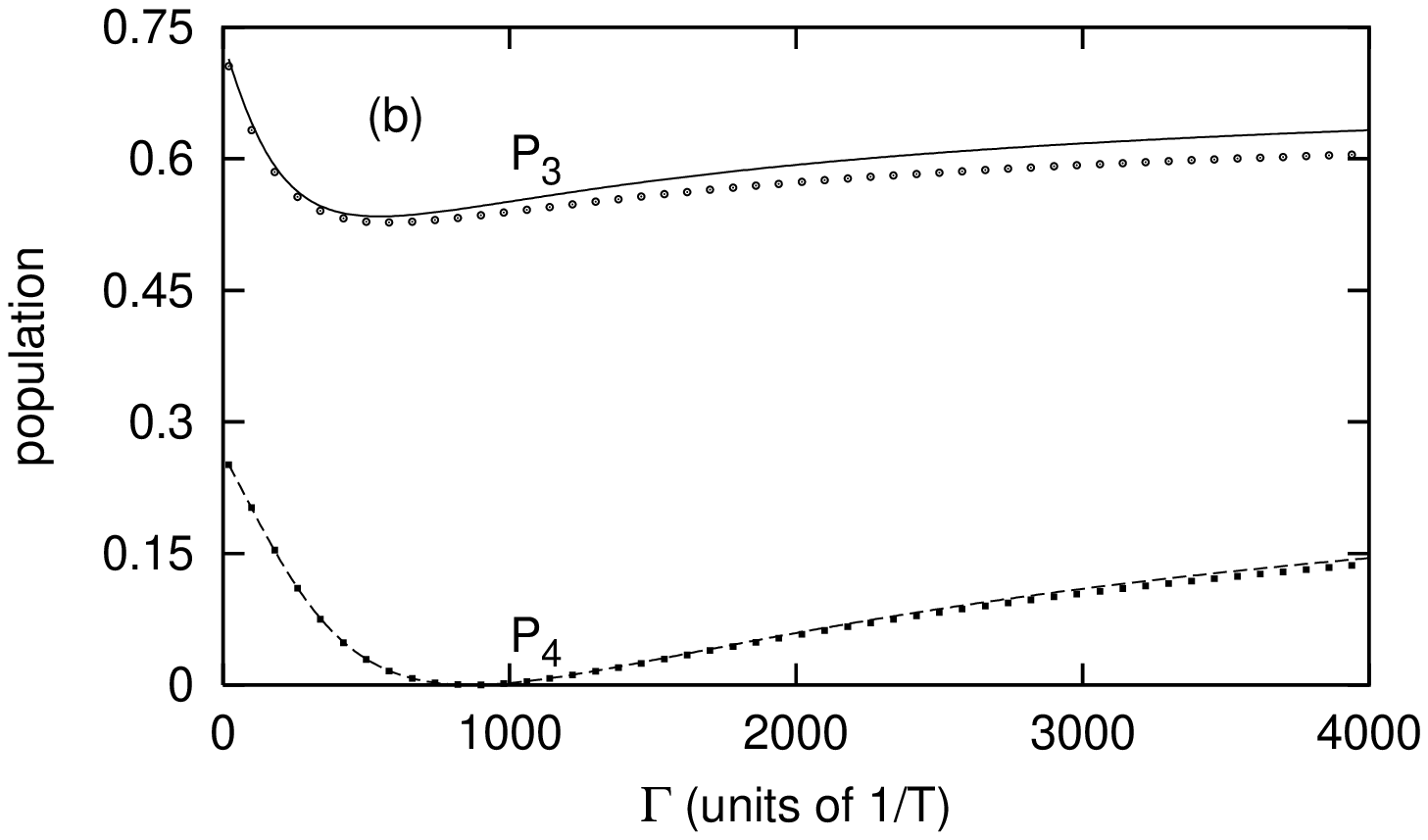,angle=270,width=6.5cm}
\end{center}
\caption{The final population in degenerate states $|3\rangle$ and  $|4\rangle$ (represented by $P_{3}$ and $P_{4}$,  $B=P_{3}/P_{4}$)
as a function of the measurement strength characterized by $\Gamma$. $\tilde{\Omega}_{P}T=10$, $\tilde{\Omega}_{B3}T=30$, $\tilde{\Omega}_{B4}T=50$. In case (a)
$\tilde{\Omega}_{S3}T=30$, $\tilde{\Omega}_{S4}T=70$ and in case (b) $\tilde{\Omega}_{S3}T=60$, $\tilde{\Omega}_{S4}T=40$.
Solid and dashed lines represent the theoretical results from Eq. (\ref{key1}), and discrete points represent
the results obtained by directly solving the Schr\"{o}dinger equation.
}
\label{fig2}
\end{figure}

Figure \ref{fig2} shows two typical computational examples of (strong) measurement assisted control of $B$. 
As seen from Fig. \ref{fig2},
the theoretical results predicted by Eq. (\ref{key1}) are in excellent agreement with those obtained by
directly solving the Schr\"{o}dinger equation. The effect of a continuous measurement of the population of the branch
state is profound. In particular, 
in case (a) the yield of state $|3\rangle$ (denoted $P_{3}$) is essentially turned off for $\Gamma T\sim 750$ whereas the yield of
state $|4\rangle$ (denoted $P_{4}$) is about $22\%$;
in case (b) $P_4$ is totally suppressed for $\Gamma T\sim 900$
whereas $P_{3}$ is as large as $53\%$. 
From Fig. \ref{fig2}a it is also seen that
even though state $|4\rangle$ is coupled more strongly
with the measured branch state than state
$|3\rangle$, $P_4$ may, counter-intuitively, be less affected by the measurement than $P_3$.
We stress that
the significant control shown in Fig. \ref{fig2} is achieved without populating the intermediate state. This is the case
because the entire two-dimensional subspace spanned by $|\lambda_{1}\rangle$ and $|\lambda_{2}'\rangle$
has a node on state $|2\rangle$, and is effectively decoupled from other adiabatic states. 
Remarkably, our further calculations show that one may also achieve $B\approx 0$ or  $B\approx \infty$ even when the natural lifetimes
of the product states are much shorter ({\it e.g.}, $T/5$) than the pulse duration \cite{gong}.

As an aside we note that $H_{\Gamma}$ can be thought of as
an ``imaginary detuning'' from the branch state. Thus, the above formalism with slight modifications ({\it e.g.}, with a real $\lambda_{2}'$)
seems to suggest that the quantum interference between states $|\lambda_{1}\rangle$ and $|\lambda_{2}'\rangle$ may also be manipulated if
we detune the branching pulse by a very large amount.
However, a large detuning often induces resonant or nearly resonant transitions to other states that are not
included in the five-level system. Moreover, 
due to the different nature of the associated nonadiabatic dynamics,
the complete suppression of one product channel based on detuning is found to be much less common
({\it e.g.}, it does not occur
in cases (a) or (b) in Fig. \ref{fig2}), and if obtained
has a far more limited tolerance to variation of the natural lifetimes of the product states
than that achieved by a strong measurement.

\section{Weak Measurement Limit}
We now turn to the case of weak system-probe interaction, {\it i.e.}, $\Gamma<<\tilde{\Omega}$.  
Here $H_{\Gamma}$ 
can be treated as a perturbation to $H_{r}$.
The null eigenvalue $\lambda_{1}$
and the null eigenvector $|\lambda_{1}\rangle$ still exist.
 Let the other eigenvalues and eigenvectors be $\lambda_{k}''$ and
$|\lambda_{k}''\rangle$, $k=2-5$.   
To the zeroth order of $H_{\Gamma}$ one has
$|\lambda_{k}''\rangle = |\lambda_{k}\rangle$.
This yields, to the first order of $H_{\Gamma}$,
\beq
\Re(\lambda_{k}'')&=&\Re(\lambda_{k});  \\
\Im(\lambda_{k}'')&=& -\Gamma \lambda_{k}^{2}(\Omega_{B3}\Omega_{S3}+\Omega_{B4}\Omega_{S4})^{2}/N_{k}^{2}.
\label{imeq}
\eeq
The above perturbative treatment gives $|\lambda_{k}''-\lambda_{1}|\approx
|\lambda_{k}-\lambda_{1}|$,
so the coherent population transfer 
can adiabatically follow $|\lambda_{1}\rangle$, $B$ therefore equals $B_{1}$, and
the measurement at first glance seems to play no role.
Nevertheless, the imaginary part of $\lambda_{k}''$ suggests that even if the undesired adiabatic states
$|\lambda_{k}''\rangle$ become populated due to some uncontrollable factors, such as the instability in laser phases 
or energy level fluctuations,
they may still be absorbed away
by the measurement which however keeps the population in the null eigenstate intact.
More significantly, we find a strong correlation between $\Im(\lambda_{k}'')$ and 
$D_{B}$, where $D_{B}$
represents the difference between $B_{1}$ and the time-evolving branching ratio given by 
$|\lambda_{k}''\rangle$. Using Eq. (\ref{eigenk}) 
one finds
\beq
D_{B}=\frac{(\lambda_{k}^{2}\Omega_{S3}-\Omega_{B4}\Omega_{SB})^{2}}{
(\lambda_{k}^{2}\Omega_{S4}+\Omega_{B3}\Omega_{SB})^{2}}-\frac{\Omega_{B4}^{2}}{\Omega_{B3}^{2}}.
\label{db}
\eeq
Comparing the above result with Eq. (\ref{imeq}) one sees that, in general,
states $|\lambda_{k}''\rangle$ with a larger $\lambda_{k}^{2}$ will give a worse $B$ 
but will be absorbed by the weak measurement  more quickly \cite{note}. Interestingly, if
$\tilde{\Omega}_{S3}\tilde{\Omega}_{B3}+\tilde{\Omega}_{S4}\tilde{\Omega}_{B4}=0$, then 
$\Im(\lambda_{k}'')=0$ at all times for $k=2-5$. But this presents 
no difficulty  since Eq. (\ref{db}) will give $D_{B}=0$ as well.
That is, in the cases where states $|\lambda_{k}''\rangle$ are unaffected by the weak measurement of the
population of the branch state,
they necessarily give the right $B$.
Clearly, then, a weak continuous measurement of the population of the branch state 
can serve as a convenient and effective ``error correction'' tool in realizing adiabatic passage.

\section{Example: Selective Population Transfer in the Presence of Strong Dephasing}
\begin{figure}[ht]
\begin{center}

\epsfig{file=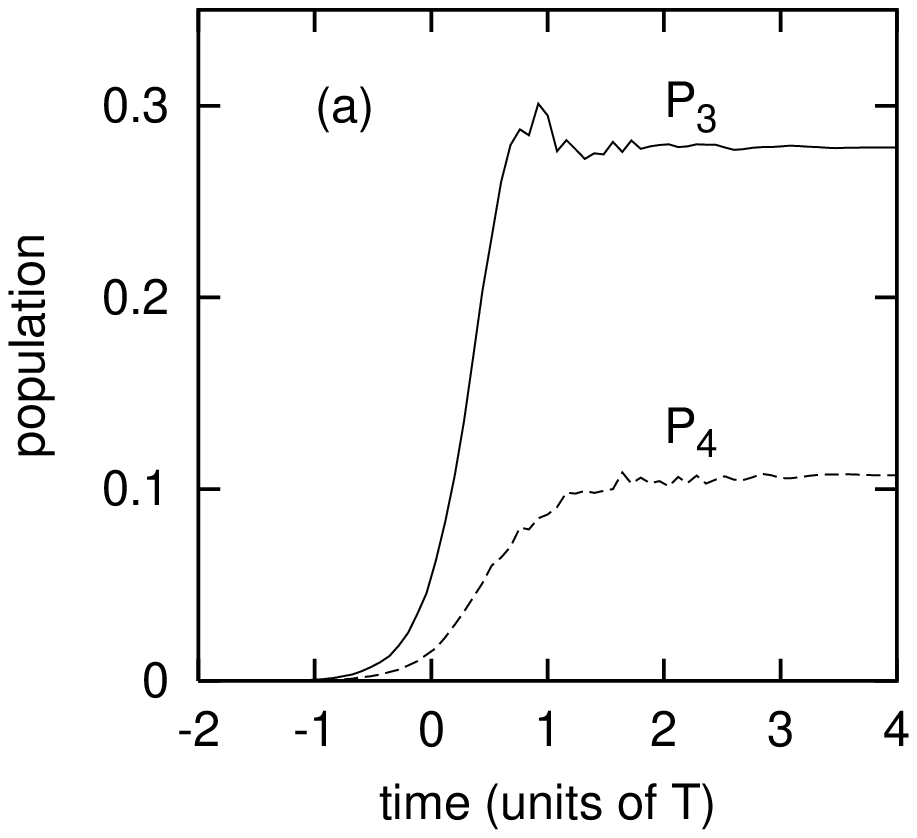,angle=270,width=3.5cm}
\epsfig{file=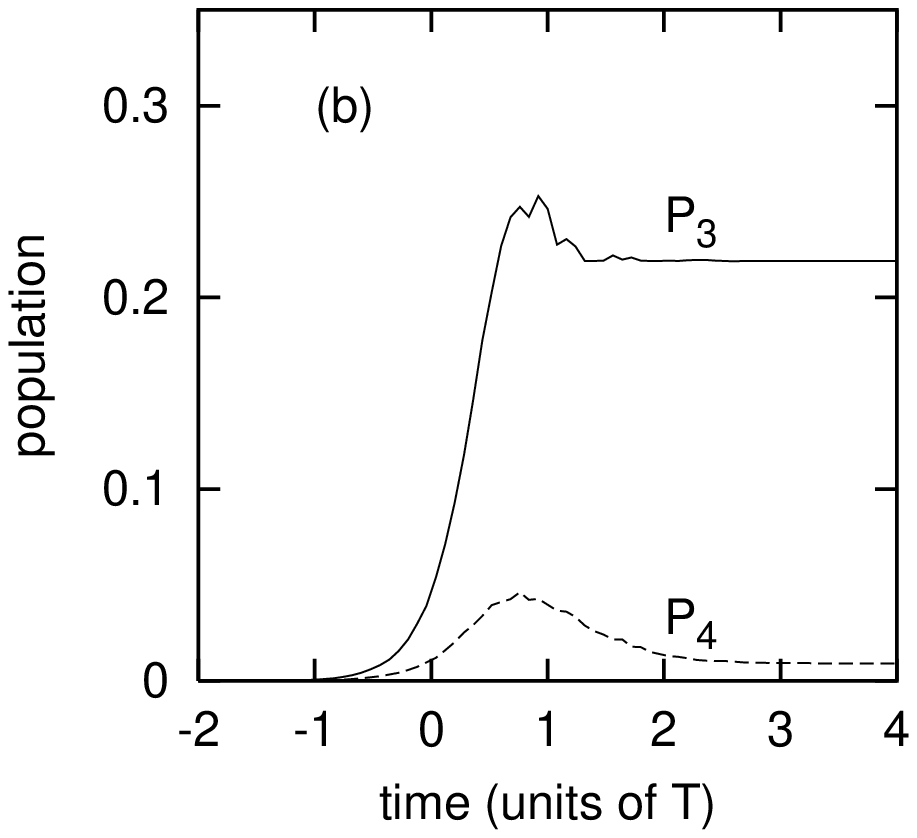,angle=270,width=3.5cm}

\epsfig{file=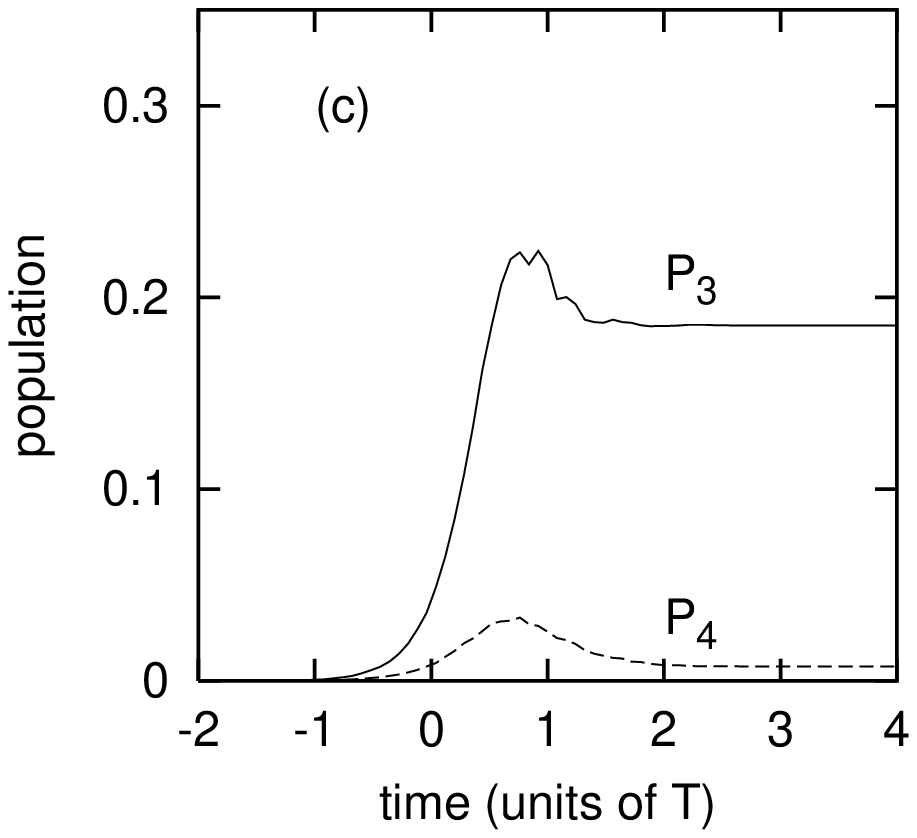,angle=270,width=3.5cm}
\epsfig{file=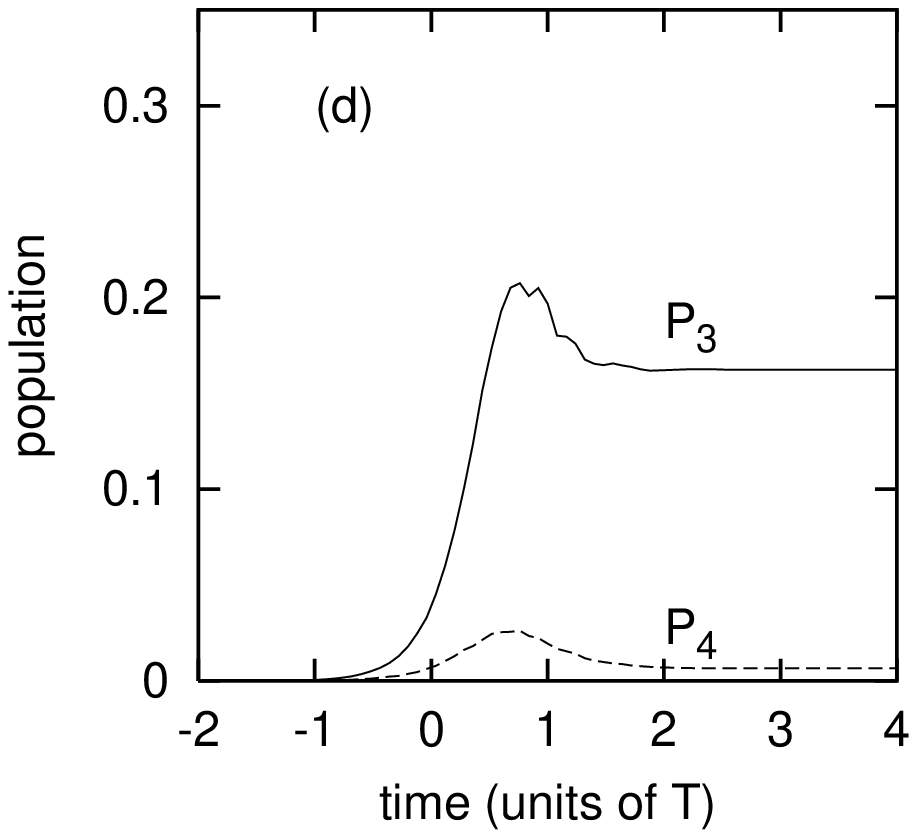,angle=270,width=3.5cm}
\end{center}
\caption{ Measurement-assisted recovery of a desired branching ratio ($B_{1}=25.0$) in the presence of strong dephasing.
Shown here is the time dependence of the population in states  $|3\rangle$ and $|4\rangle$ (represented by $P_{3}$ and $P_{4}$)
without (case a) measurement, or  with measurement for (b) $\Gamma T=3$, (c)  $\Gamma T=6$ , and (d) $\Gamma T=9$.
 $\tilde{\Omega}_{P}T=20$, $\tilde{\Omega}_{S3}T=50$, $\tilde{\Omega}_{S4}T=40$,
$\tilde{\Omega}_{B3}T=15$, $\tilde{\Omega}_{B4}T=75$. The results are obtained by averaging over 1000 realizations of the stochastic
energy level fluctuations.
}
\label{fig3}
\end{figure}

We now apply the finding of the previous section to selective population transfer in the presence of strong dephasing \cite{mustafa,demirplak,gongjcp},
where
the dephasing (caused by, e.g., the coupling with other degrees of freedom, bimolecular collisions, or solvent effects on a solute molecule) 
is simulated by
stochastic Gaussian fluctuations $\delta\omega_{k}$ ($k=1-5$) associated with each of the five levels.
The degeneracy between states $|3\rangle$ and $|4\rangle$ is assumed to be maintained, {\it i.e.}, $\delta\omega_{3}=\delta\omega_{4}$.
The mean properties of $\delta\omega_{k}$ are chosen to be
$\langle\delta\omega_{k}\rangle=0$ and
$\langle\delta\omega_{k}(t)\delta\omega_{k'}(t')\rangle=\delta_{kk'}\Delta^{2}
\exp[-|t-t'|/\tau]$, $k,k'=1-3, 5$.
A computational example
is shown in Fig. \ref{fig3},
where $\Delta T=15$ (a value that represents extremely strong dephasing)
and $\tau=0.02T$ (a value that is comparable
to $1/\tilde{\Omega}$ to further enhance the dephasing-induced  nonadiabatic effects \cite{demirplak,gongjcp}).
The goal in this example is to achieve $B=25.0$ and it is assumed that
there indeed exists a branch state that meets this goal, {\it i.e.}, $B_{1}=25.0$.
As is seen in Fig. \ref{fig3}a, the value of $B$ without the measurement is terribly degraded by dephasing and is
as small as $2.6$.
However, upon introducing a weak measurement of the population of the branch state,
we obtain a dramatic increase to $B=24.0$ for $\Gamma T=3 $ (Fig. \ref{fig3}b),  $B=24.9$ for $\Gamma T=6$  (Fig. \ref{fig3}c) and
for $\Gamma T=9$ (Fig. \ref{fig3}d).
As seen in Fig. \ref{fig3},  this almost perfect recovery of the desired value of $B$ is attained with
a considerable total yield of the products and with a wide range of the (weak) measurement strength.
We have obtained
similar results
in numerous other cases, such as those with  different $\tau$, $B_{1}$, $\tilde{\Omega}_{S3}/\tilde{\Omega}_{S4}$, or different timing and duration
of the branching pulse.

\section{Concluding Remarks}
It is important to note that although the measurement of the population of the branch state absorbs away  some population by, for
example, breaking up some molecules from an ensemble of molecules that are subject to the control fields,
our purpose in introducing it is to generate desired quantum interference effects.
In the strong measurement case considered in Secs. II and III, the measurement
creates a second adiabatic state that can be significantly populated during
the population transfer and thereby
induces quantum interference effects that are absent in the
unmeasured five-level STIRAP
system. 
In the weak measurement case considered in Secs. IV and V, the 
measurement of the population of the branch state
enhances the robustness
of the quantum
interference effects inherent in the unmeasured five-level STIRAP system. 
That is,  during the population transfer
those molecules that are broken up by the weak measurement have the potential
to generate
a wrong branching ratio, 
and those that survive the weak measurement carry the desired quantum 
interference
induced by the strong control fields.

Hence, while a measurement on a quantum system subject to coherent control
necessarily introduces decoherence to the system, it
can still be beneficial for coherent control,  with the price that the
total yield of the product states is less than 100\%.
Nevertheless, one should never let the best be the enemy of the good.
 This price is acceptable,
particularly when the main objective of coherent control
is to alter at will the branching ratio
between two product states.  
The results of our study can also be regarded as an extension of our previous work \cite{gongjcp}, where we have shown that
the decay of a target state or measuring  the population of the target state can greatly improve the performance
of a three-level STIRAP system in the presence of strong dephasing \cite{mustafa,yat,geva}.

In conclusion, we find that the  population transfer branching ratio
between degenerate
product states 
may be significantly
altered by observing the system.  Our results
suggest that coherent control can be even more powerful
than previously thought if we combine
quantum interference effects with  measurement effects.
The effects we have described depend on the character of the influences induced when the population
of the branch state is measured, and are different from the kind of final state branching ratio that
develops as the result of incoherent kinetic competition between parallel pathways.
Potential applications of this work include control of molecular chirality,
control of photodissociation, and
control of isomerization reactions
in solution. 

\vspace{0.5cm}
\section{acknowledgments}
This work was supported by the National Science Foundation.


\begin{thebibliography}{100}
\bibitem{ricebook}
 S.A. Rice and M. Zhao, {\it Optical Control of Molecular Dynamics} (John
  Wiley, New York, 2000).
\bibitem{rice01} S.A. Rice, Nature {\bf 409}, 422 (2001).
\bibitem{brumerbook}
 M. Shapiro and P. Brumer, {\it Principles of
 the Quantum Control of Molecular Processes} (John Wiley, New York, 2003).
\bibitem{bergmann}K. Bergmann, H. Theuer, and B.W. Shore, \rmp{\bf 70}, 1003 (1998).
\bibitem{rabitz}
H. Rabitz, R. de Vivie-Riedle, M. Motzkus, and K. Kompa,
 Science {\bf 288},
824 (2000).
\bibitem{vera} V. Frerichs and A. Schenzle, \pra{\bf 44}, 1962 (1991).
\bibitem{schulman} L.S. Schulman, \pra{\bf 57}, 1509 (1998).
\bibitem{luis} A. Luis, \prl{\bf 76}, 4340 (1996); \pra{\bf 67}, 062113 (2003).
\bibitem{facchi2} P. Facchi and S. Pascazio, Fortschr. Phys. 
{\bf  49}, 941 (2001); \prl{\bf 89}, 080401 (2002). 
\bibitem{kobrakpra} M.N. Kobrak and S.A. Rice, \pra{\bf 57}, 2885 (1998);
\jcp{\bf 109}, 1 (1998).
\bibitem{kurkal}V. Kurkal and S.A. Rice, J. Phys. Chem. B {\bf 105}, 6488 (2001). 
\bibitem{note2} 
It is exactly the resonant laser-molecule coupling that makes it possible to
describe a molecular system in terms of a few energy levels. For example,
Ref. \cite{kurkal} has shown that the Kobrak-Rice five-level model works well in the presence
of nonresonant coupling with a large number of background states.
\bibitem{gong}J. Gong and S.A. Rice, unpublished.
\bibitem{note} Although measuring the intermediate state $|2\rangle$ will also
keep $|\lambda_{1}\rangle$
intact and introduce nonzero $\Im(\lambda_{k}'')$,
it is not useful since the strong correlation between $D_{B}$ and $\Im(\lambda_{k}'')$
no longer exists. 
\bibitem{mustafa}M. Demirplak and S.A. Rice, \jcp{\bf 116}, 8028 (2002).
\bibitem{demirplak}M. Demirplak and S.A.  Rice, J. Phys. Chem. A {\bf 107}, 9937 (2003).
\bibitem{gongjcp} J. Gong and S.A. Rice, J. Chem. Phys. {\bf 120}, 3777 (2004).
\bibitem{yat} L.P. Yatsenko, V.I. Romanenko, B. W. Shore, and K. Bergmann, \pra{\bf 65}, 043409 (2002).
\bibitem{geva} Q. Shi and E. Geva, \jcp{\bf 119}, 11773 (2003).

\end{thebibliography}
\end{document}